\documentclass[10pt]{article}
\usepackage{amsmath}
\usepackage{epsfig}
\usepackage{url}
\newlength{\figwidth}
\ifdefined\jinst
\newcommand{\VECTOR}[1]{{\ensuremath{\boldsymbol{\mathsf{#1}}}}}
\else
\newcommand{\VECTOR}[1]{{\ensuremath{\boldsymbol{#1}}}}
\fi
\newcommand{\MATRIX}[1]{{\ensuremath{\boldsymbol{\mathsf{#1}}}}}
\newcommand{\TR}[1]{\ensuremath{#1^{\sf T}}}
\newcommand{\MATRIXTR}[1]{\TR{\MATRIX{#1}}}
\newcommand{\VECTORTR}[1]{\TR{\VECTOR{#1}}}
\newcommand{\mytitle}{TUnfold, an algorithm for correcting migration effects in high energy physics}
\newcommand{\myauthor}{Stefan Schmitt}
\newcommand{\myinstitute}{DESY, Notkestra\ss{}e 85, 22607 Hamburg}
\newcommand{\myemail}{email: sschmitt@mail.desy.de}
\newcommand{\mydate}{September 2012}
\newcommand{\myabstract}{
TUnfold is a tool for correcting migration and background effects in
high energy physics for multi-dimensional distributions.
It is based on a least square fit with Tikhonov
regularisation and an optional area constraint. For determining the
strength of the regularisation parameter, the L-curve method and scans of
global correlation coefficients are 
implemented. The algorithm supports background subtraction and the 
propagation of statistical and systematic uncertainties, in particular 
those originating from limited knowledge of the response matrix.
The program is interfaced to the ROOT analysis framework.
}
\ifdefined\jinst
\title{\mytitle}
\author{\myauthor\\ \myinstitute\\ E-mail:\email{\myemail}}
\abstract{\myabstract}
\begin{document}
\else
\ifdefined\institute
%
% EPJ
\begin{document}
\title{\mytitle}
\author{\myauthor\thanks{\myemail}}
\institute{\myinstitute}
\date{\mydate}
\abstract{
\myabstract}
\maketitle
%
% DESY preprint format
\else
\renewcommand{\topfraction}{1.0}
\renewcommand{\bottomfraction}{1.0}
\renewcommand{\textfraction}{0.0}
\newlength{\dinwidth}
\newlength{\dinmargin}
\setlength{\dinwidth}{21.0cm}
\textheight23.5cm \textwidth16.0cm
\setlength{\dinmargin}{\dinwidth}
\setlength{\unitlength}{1mm}
\addtolength{\dinmargin}{-\textwidth}
\setlength{\dinmargin}{0.5\dinmargin}
\oddsidemargin -1.0in
\addtolength{\oddsidemargin}{\dinmargin}
\setlength{\evensidemargin}{\oddsidemargin}
\setlength{\marginparwidth}{0.9\dinmargin}
\marginparsep 8pt \marginparpush 5pt
\topmargin -42pt
\headheight 12pt
\headsep 30pt \footskip 24pt
\parskip 3mm plus 2mm minus 2mm
\begin{document}
%
% DESY titlepage
\begin{titlepage}
\begin{flushleft}
{\tt DESY 12-129    \hfill    ISSN 0418-9833} \\
{\tt \mydate}                  \\
\end{flushleft}

\vspace{2cm}
\begin{center}
\begin{Large}
{\bf\mytitle} \\
\vspace{2cm}

\myauthor, \myinstitute\\
\myemail
\end{Large}
\end{center}

\vspace{2cm}
\begin{abstract}
\myabstract
\end{abstract}
\vspace{1.5cm}

\begin{center}
Accepted by JINST as technical report
\end{center}
\end{titlepage}
\fi\fi

\section{Introduction}

In high energy physics, experiments are usually performed as counting
experiments, where events are grouped into certain regions of
phase-space, also called bins. However, the kinematic properties of
each event, such as four-momenta of particles and derived quantities,
are measured only at finite precision due to inevitable detector
effects.
As a consequence, events may be found in the wrong bin.
Furthermore there is the presence of background, such that
only a fraction of the events observed in a given bin originates from
the reaction one is interested in.

In most cases, algorithms such as GEANT \cite{geant} are
used to simulate migrations imposed by detector effects, whereas
underlying physics processes are simulated using event generators such
as PYTHIA \cite{Sjostrand:2000wi}.
After tracking the generated events through the
detector simulation one is able to confront the physics process
modelled by the event generator with the background-subtracted data.

\begin{figure}[ht]
\begin{center}
\includegraphics[width=0.5\figwidth]{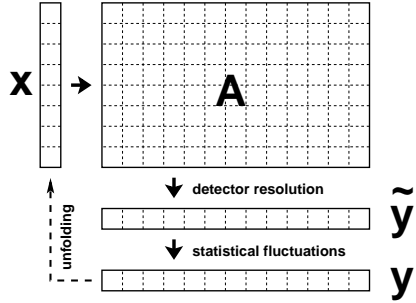}
\caption{\label{fig:simplescheme}schematic view of migration effects and
  statistical fluctuations}
\end{center}
\end{figure}

However, often one is interested to report results such as
differential cross sections, independent of the detector
simulation. In that case, the observed event counts have to be
corrected for detector effects. The problem may be written as
\begin{equation}
\tilde{y}_i = \sum_{j=1}^m A_{ij}\tilde{x}_j,\,1\le i\le n\label{eqn:basic}
\end{equation}
where the $m$ bins $\tilde{x}_j$ represent the true distribution, $A_{ij}$ is
a matrix of probabilities describing the migrations from
bin $j$ to any of the $n$ bins on detector level and $\tilde{y}_i$ is
the average expected event count at detector level.
It is important to note here that the observed event counts $y_i$ may
be different from the average $\tilde{y}_i$ due to statistical fluctuations.
A schematic view is given in figure \ref{fig:simplescheme}.
The situations becomes somewhat more complicated if there is
background. In that case the $\tilde{y}_i$ receive an additional
contribution from background,
\begin{equation}
\tilde{y}_i = \sum_{j=1}^m A_{ij}\tilde{x}_j+b_i,\,1\le i\le n\label{eqn:basicbgr}
\end{equation}
where $b_{i}$ is the background showing up in bin $i$. Both the
background and the matrix of probabilities often suffer from systematic
uncertainties which have to be considered in addition to the
statistical uncertainties.

One may be tempted to replace $\tilde{y}_i\to y_i$ and $\tilde{x}_j\to x_j$
in equations \ref{eqn:basic} or \ref{eqn:basicbgr} and then solve
for $x_j$, simply by inverting the matrix of probabilities. However,
it turns out that the statistical fluctuations of the $y_i$ are
amplified when calculating the $x_j$ this way. Such fluctuations are
often damped by imposing certain smoothness conditions on the
$x_j$. This procedure is termed ``regularisation''.

The TUnfold algorithm \cite{tunfold17}, described in this
paper and interfaced to the ROOT analysis package \cite{Brun:1997pa},
implements a procedure to estimate the $\tilde{x}_j$ using a least square method
with Tikhonov regularisation \cite{tikhonov} and an optional area constraint.
In order to obtain best results from the least square minimisation,
the number of degrees of freedom, $n-m$, has to be larger than
zero. It means that the data $y_i$ have to be measured in finer bins
than are extracted by the unfolding procedure.
This condition $n\ge m$ is in contrast to some other commonly used
unfolding methods, where often the restriction $n=m$ is imposed
\cite{Hocker:1995kb,D'Agostini:1994zf}. Examples of unfolding
algorithms which do not have the restriction $n=m$ are
\cite{D'Agostini:new,Choudalakis:2012hz}.

No attempt is made here to give a complete overview of the 
commonly used unfolding algorithms. The TUnfold algorithm
\cite{tunfold17} presented here compares best to algorithms based on
matrix inversion or singular value decomposition, like
\cite{Hocker:1995kb,Blobel:2002pu}. Alternative approaches are often
based on iterative methods or on the use of Bayes' theorem, for
example \cite{D'Agostini:1994zf,D'Agostini:new,Choudalakis:2012hz}.
Many reviews on the topic can be found in literature,
only two examples are given here \cite{Anykeev:1990qe,phystat2011}.

\section{The TUnfold algorithm}

\subsection{Definitions}

The TUnfold algorithm gives an estimator of a set of truth parameters,
using a single measurement of a set of observables.
The observables are described by a vector\footnote{Throughout this
  paper, matrices (\MATRIX{M}) and vectors (\VECTOR{v}) are printed in bold.
  Matrices or vectors without indices, written next
  to each other, are multiplied. Where needed, brackets with indices
  are used to refer to specific elements.
  The notation $\MATRIXTR{M}$ indicates that a matrix is
  transposed, its rows and columns are swapped.
  The inverse of $\MATRIX{M}$ is written
  as $\MATRIX{M}^{-1}$. A vector is treated as a
  matrix with only one column, such that a transposed vector has only one row.
  The dot product of two vectors $\VECTOR{v_1}$ and $\VECTOR{v_2}$
  thus is equivalent to the
  matrix multiplication $\VECTORTR{v_1}\VECTOR{v_2}$. Other examples are
  $(\MATRIX{A}\VECTOR{x})_i=\sum_j A_{ij}x_j$ and
  $(\MATRIXTR{A})_{ij}=A_{ji}$.}
of random variables, $\VECTOR{y}$. The random variables $\VECTOR{y}$
are taken to have a multivariant Gaussian distribution with mean
$\tilde{\VECTOR{y}}=\MATRIX{A}\tilde{\VECTOR{x}}$, where
$\tilde{\VECTOR{x}}$ is a vector corresponding to the set of of truth
parameters and $\MATRIX{A}$ is a matrix.
The covariance matrix of $\VECTOR{y}$ is
$\MATRIX{V_{yy}}$. The algorithm only works if the dimension of
$\tilde{\VECTOR{x}}$ is less or equal to the dimension of
$\tilde{\VECTOR{y}}$. Furthermore, $\MATRIX{V_{yy}}$ has to have full
rank and the rows of $\MATRIX{A}$ shall be linear independent.
The algorithm returns an estimator $\VECTOR{x}$ of the truth
parameters $\tilde{\VECTOR{x}}$, given an observation $\VECTOR{y}$.
The estimator $\VECTOR{x}$, when considered as a random variable, has a
covariance matrix which is also calculated. It is labelled
$\MATRIX{V_{xx}}$.

\subsection{Algorithm}
 
The unfolding algorithm, as implemented in TUnfold, determines the stationary
point of the ``Lagrangian''
\begin{align}
{\cal L}(x,\lambda) = & {\cal L}_1 + {\cal L}_2 + {\cal L}_3
\label{eqn:lagrangian} & \text{where} \\
{\cal L}_1 = & \TR{(\VECTOR{y}-\MATRIX{A}\VECTOR{x})}
\MATRIX{V_{yy}}^{-1}
(\VECTOR{y}-\MATRIX{A}\VECTOR{x}), & \label{eqn:leastsquare} \\
{\cal L}_2 = & \tau^2
\TR{(\VECTOR{x}-f_b \VECTOR{x_0})}(\MATRIXTR{L}\MATRIX{L})(\VECTOR{x}-f_b
\VECTOR{x_0}), & \label{eqn:regularisation} \\
{\cal L}_3 = & \lambda (Y-\VECTORTR{e}\VECTOR{x})\label{eqn:constraint} & \text{and} \\
Y = & \sum_i y_i\label{eqn:datatotal}, & \\
e_j = & \sum_i A_{ij}, &\label{eqn:efficiency}
\end{align}

The term ${\cal L}_1$ is what one expects from a least square minimisation.
The vector $\VECTOR{y}$ has $n$ rows. The covariance matrix $\MATRIX{V_{yy}}$ of
$\VECTOR{y}$ is diagonal in many cases, such that the diagonals hold
the squares of the uncertainties. TUnfold also supports the use of non-diagonal
$\MATRIX{V_{yy}}$.
The vector $\VECTOR{x}$ corresponds to the unfolding result and has $m$ rows.
The elements $A_{ij}$ of $\MATRIX{A}$ describe for each row $j$ of
$\VECTOR{x}$ the probabilities to migrate to bin $i$ of $\VECTOR{y}$. The
matrix $\MATRIX{A}$ often is determined using Monte Carlo simulations.

The term ${\cal L}_2$ describes the regularisation, which damps
fluctuations in $\VECTOR{x}$. Such fluctuations originate from the
statistical fluctuations of $\VECTOR{y}$, which are amplified when determining
the stationary point of equation \ref{eqn:lagrangian}. The parameter
$\tau^2$ gives the strength of the regularisation. It is considered as
a constant while determining the stationary point of ${\cal L}$.
The matrix $\MATRIX{L}$ has $n$ columns and $n_R$ rows,
corresponding to $n_R$ regularisation conditions. 
The bias vector $f_b \VECTOR{x_0}$ is composed of a
normalisation factor $f_b$ and a vector $\VECTOR{x_0}$. 
In the simplest case, one has $f_b=0$,  $n_R=n$ and $\MATRIX{L}$ is
the unity matrix. In that case, ${\cal L}_2$ simplifies to $\tau^2
\vert\vert \VECTOR{x}\vert\vert^2$, effectively suppressing large
deviations of \VECTOR{x} from zero. If $f_b=1$, deviations of
\VECTOR{x} from \VECTOR{x_0} are suppressed. Choices of
the matrix \MATRIX{L} different from the unity matrix are discussed in section \ref{text:regularisation}.

The term ${\cal L}_3$ is an optional area
constraint. There is a Lagrangian parameter $\lambda$. The sum over
all observations is given by $Y$, equation~\ref{eqn:datatotal}.
The efficiency vector $\VECTOR{e}$ has $m$ rows and is calculated
from $\MATRIX{A}$ as indicated in equation~\ref{eqn:efficiency}.
If the area constraint is used, the normalisation of the result
$\VECTOR{x}$, corrected for the efficiencies $\VECTOR{e}$, is thus
enforced to match the total event count $Y$.
This procedure is applied in order to limit possible biases on the
normalisation which are present if the data $\VECTOR{y}$ follow
Poisson's statistics whereas the least square ansatz is strictly valid only
for normal distributed measurements. The problem is discussed in more detail
in literature, for example in \cite{Cowan:1998}.

The minimum or stationary point of ${\cal L}(\VECTOR{x},\lambda)$ is
determined by setting the first derivatives to zero. In
the case without area constraint, $\lambda$ is set to zero and only the
derivatives of  ${\cal L}_1+{\cal L}_2$ with respect to the components of
\VECTOR{x} are set to zero.
When including the area constraint, the equations are solved for
\VECTOR{x} and $\lambda$ together. The partial derivatives of ${\cal
  L}(\VECTOR{x},\lambda)$ are
\begin{align}
\frac{\partial {\cal L}(\VECTOR{x},\lambda)}{\partial x_j} = & -2\left(\MATRIXTR{A}\MATRIX{V_{yy}}^{-1}(\VECTOR{y}-\MATRIX{A}\VECTOR{x})\right)_j+ 2\tau^2\left((\MATRIXTR{L}\MATRIX{L})(\VECTOR{x}-f_b
\VECTOR{x_0})\right)_j - \lambda e_j, & \\
\frac{\partial {\cal L}(\VECTOR{x},\lambda)}{\partial\lambda} = & Y-\VECTORTR{e}\VECTOR{x}. & \\
\end{align}
The stationary point $\VECTOR{x}$ of ${\cal L}$ is found as
\begin{align}
 \VECTOR{x} = & \begin{cases} 
  \VECTOR{x\vert_{\lambda=0}} & \text{without area constraint} \\
  \VECTOR{x\vert_{\lambda=0}} + \frac{\lambda}{2} \MATRIX{E}\VECTOR{e} & \text{with area
   constraint}\end{cases} & \text{where} \\
\VECTOR{x\vert_{\lambda=0}} = & \MATRIX{E}\left[\MATRIXTR{A}\MATRIX{V_{yy}}^{-1}\VECTOR{y}
  +\tau^2(\MATRIXTR{L}\MATRIX{L})f_b \VECTOR{x_0}\right], & \\
 \MATRIX{E} = & \left(\MATRIXTR{A}\MATRIX{V_{yy}}^{-1}\MATRIX{A}+ \tau^2(\MATRIXTR{L}\MATRIX{L})\right)^{-1} & \text{and} \\
 \frac{\lambda}{2} = & \frac{Y -\VECTORTR{e} \VECTOR{x\vert_{\lambda=0}}}{\VECTORTR{e}\MATRIX{E}\VECTOR{e}}. &
\end{align}
In order to calculate the covariance matrix of $\VECTOR{x}$, given
the covariance matrix of $\VECTOR{y}$, the
corresponding partial derivatives are calculated
\begin{align}
(\MATRIX{D^{xy}})_{ki} : = \frac{\partial x_k}{\partial y_i} = & \begin{cases}
B_{ki} & \text{without area constraint} \\
B_{ki} +
(\MATRIX{E}\VECTOR{e})_k\frac{1-(\MATRIXTR{B}\VECTOR{e})_i}{\VECTORTR{e}\MATRIX{E}\VECTOR{e}} & 
\text{with area constraint} \end{cases} &
\text{where} \\
\MATRIX{B} = & \MATRIX{E}\MATRIXTR{A}\MATRIX{V_{yy}}^{-1}. &
\end{align}
The covariance matrix of the result $\VECTOR{x}$, originating from
$\MATRIX{V_{yy}}$ is thus given by
\begin{equation}
\MATRIX{V_{xx}} = \MATRIX{D^{xy}}\MATRIX{V_{yy}} \TR{(\MATRIX{D^{xy}})}.
\end{equation}

\section{Normalisation of the matrix of migrations}

In most cases, $\MATRIX{A}$ is determined from Monte Carlo
simulations. Within TUnfold, it is foreseen to initialise the unfolding
from a matrix $\MATRIX{M}$ of event counts, determined in a Monte Carlo event
simulation, where $\MATRIX{M}$ has $n+1$
rows and $m$ columns, one row more than $\MATRIX{A}$.
The extra row is used to count those events which are generated in
a particular bin $j$ but are not found in any of the reconstructed bins.
For the purpose of this paper, the extra row of $M$ is denoted with index
$i=0$, whereas all other matrices and vectors have indices starting
from $1$. In other words, the matrix elements $M_{ij}$ count the Monte
Carlo events generated in bin $j$ of $\VECTOR{x}$ and reconstructed in bin
$i>0$ of $\VECTOR{y}$, whereas the matrix elements $M_{0j}$ count the
Monte Carlo events generated in bin $j$ and not reconstructed in any
of the bins of $\VECTOR{y}$.
For the unfolding algorithm, 
$\MATRIX{A}$ and $\VECTOR{x_0}$ are initialised from $\MATRIX{M}$ as follows
\begin{align}
A_{ij} = & \frac{M_{ij}}{s_j},\,\text{where }i>0 & \text{ and}
\label{eqn:MtoA} \\
s_j = & \sum_{i=0}^n M_{ij}, & \\
(\VECTOR{x_0})_j = & s_j. &
\end{align}

\section{Choice of the regularisation strength}

When unfolding, the strength of the regularisation, $\tau^2$, is an
unknown parameter. If $\tau^2$ is too small, the unfolding result
often has large fluctuations and correspondingly large negative
correlations of adjacent bins. If $\tau^2$ is too large, the result is
biased towards $f_b \VECTOR{x_0}$.
Several methods to choose the strength of the regularisation are
discussed in literature, for example eigenvalue analyses \cite{eigenvalueanalysis}, minimisation
of correlation coefficients  \cite{globalcorr}, and the L-curve method
\cite{hansen2000}.
At present, in TUnfold a simple version of the L-curve method is
implemented to determine $\tau^2$ as well as methods to minimise
global correlation coefficients.

\subsection{L-curve scan}
The idea of the L-curve method is to look at the graph of two
variables $L^{\text{curve}}_x$ and $L^{\text{curve}}_y$ and locate the
point where the curvature is maximal.
These variables are defined as
\begin{align}
L^{\text{curve}}_x = & \log {\cal L}_1  & \text{and} \\
L^{\text{curve}}_y = & \log \frac{{\cal L}_2}{\tau^2}, &
\end{align}
such that $L_x$ tests the agreement of $x$ with the data and $L_y$
tests the agreement of $x$ with the regularisation condition.
Note that $L^{\text{curve}}_y$ does not have an explicit dependence on $\tau^2$.
For $\tau^2\to 0$ the value of $L^{\text{curve}}_x$ is minimal and
$L^{\text{curve}}_y$ is maximal, because ${\cal L}_2\to 0$ and $\VECTOR{x}$
corresponds to the stationary point of ${\cal L}_1+{\cal L}_3$.
As $\tau^2$ gets large, $L^{\text{curve}}_x$ increases
whereas $L^{\text{curve}}_y$ is getting small, because the Lagrangian is
dominated by ${\cal L}_2$.
It is observed that the parametric plot of $L^{\text{curve}}_y$
against $L^{\text{curve}}_x$ often shows a kink (is {\sf L}-shaped). The
kink location is chosen to determine $\tau^2$.

In TUnfold, the L-curve algorithm is implemented as follows: the
unfolding is repeated for a number of points in $t=\log\tau$, thus
scanning the L-curve. The curvature $\cal C$ of the L-curve is
determined as
\begin{equation} 
{\cal C} =
 \frac{\mathrm{d}^2L^{\text{curve}}_y\mathrm{d}L^{\text{curve}}_x-
 \mathrm{d}^2L^{\text{curve}}_x\mathrm{d}L^{\text{curve}}_y}{
\left((\mathrm{d}L^{\text{curve}}_x)^2+(\mathrm{d}L^{\text{curve}}_y)^2\right)^\frac{3}{2}
}.
\end{equation}
The first and second derivatives of $L^{\text{curve}}_x$
($L^{\text{curve}}_y$) with respect
to $t$, $\mathrm{d}L^{\text{curve}}_{x}$
($\mathrm{d}L^{\text{curve}}_{y}$) and
$\mathrm{d}^2L^{\text{curve}}_{x}$
($\mathrm{d}^2L^{\text{curve}}_{y}$), respectively, are approximated 
using cubic spline parametrisations of the scan results.
The maximum of ${\cal
  C}$ is finally determined with the help of a cubic spline
parametrisation of ${\cal C}(t)$.

\subsection{Minimising global correlation coefficients}

A method of minimising global correlation coefficients is also
implemented.
Given the covariance matrix $\MATRIX{V_{xx}}$
the global correlation coefficient of a component $i$ of $\VECTOR{x}$ is
defined as
\begin{equation}
\rho_i = \sqrt{1-\frac{1}{(\MATRIX{V_{xx}}^{-1})_{ii}(\MATRIX{V_{xx}})_{ii}}}.
\end{equation}
Two sorts of correlation coefficients scans have been implemented:
\begin{enumerate}
\item minimising the average correlation: the regularisation
  strength $\tau^2$ is chosen such that the average global
  correlation $\sum_i\rho_i/n$ is minimised, where $n$ is the dimension
  of $\VECTOR{x}$.
\item minimising the maximum correlation: the regularisation strength
  $\tau^2$ is chosen such that the maximum correlation
  $\max_i(\rho_i)$ is minimised.
\end{enumerate}

Furthermore, it is possible to choose the covariances
\begin{enumerate}
\item The covariance matrix $V_{xx}$ may or may not include systematic
  uncertainties.
\item There is the option to partition the covariance matrix such that
  only parts of the matrix are used for the calculation of global
  correlation coefficients.
\item It is possible to merge bins or groups of bins prior to calculating
  the $\rho_i$.
\end{enumerate}
When partitioning the covariance matrix, the corresponding unused
rows and columns of  $\MATRIX{V_{xx}}$ are removed prior to inverting
the matrix and calculating the global correlation coefficients. When
merging bins of groups of bins, the corresponding rows or columns of
the matrix are added up.

The scan is implemented such that the unfolding is repeated for a
number of points in $t=\log\tau$. For each point the chosen correlation
type (maximum or average) is calculated. The minimum is determined using a
cubic spline interpolation.

\section{Background subtraction}

Often there is background present in the measured data $\VECTOR{y}$. 
It is worth to mention that the background has to include all types of
events which are possibly reconstructed in one of the bins of $\VECTOR{y}$ 
but do not originate from any of the bins of $\VECTOR{x}$. In
particular, part of the signal process may be generated outside the
phase-space covered by $\VECTOR{x}$ and thus has to be counted as
background.
Sometimes it is possible to determine background sources from the
data as a part of the unfolding process, for example using a discriminator
\cite{Aaron:2010uj}. In order to achieve that, background
normalisation factors are included as extra bins of the vector
\VECTOR{x}, corresponding to extra columns of the matrices \MATRIX{A}, \MATRIX{M}. The 
background normalisation is then 
determined in the unfolding process.

On the other hand, it is often useful to simply
subtract the background prior to unfolding. Within TUnfold, the
following method of background subtraction is implemented
\begin{align}
\VECTOR{y} = & \VECTOR{y^0} - f^b \VECTOR{b}, & \\
(\MATRIX{V_{yy}})_{ij} = & (\MATRIX{V^0_{yy}})_{ij} +
\delta_{ij}(f^b(\VECTOR{\delta b})_i)^2
  +(\delta\! f^b)^2 b_i b_j. &
\end{align}
Here, the components of $\VECTOR{y^0}$ are the data
prior to background subtraction, with 
covariance matrix $\MATRIX{V^0_{yy}}$.
The background has a normalisation
factor $f^b$ with uncertainty $\delta\! f^b$. The background shape is
described by a vector $\VECTOR{b}$ and the uncertainties on the components of
$\VECTOR{b}$ are given by the vector $\VECTOR{\delta b}$.
Finally, $\delta_{ij}$ is the Kronecker symbol.

The covariance matrix $\MATRIX{V_{yy}}$ receives contributions from
the covariance matrix of \VECTOR{y^0} as well as from the
uncertainties on the background shape, the latter contributing only to
the diagonal elements. In addition there are contributions to the
covariance matrix 
from the background normalisation uncertainty. Because the background normalisation is 
correlated for all analysis bins, it also contributes to the off-diagonal
elements of the matrix.

In TUnfold, the background subtraction is generalised such that multiple
background sources may be subtracted. The contribution of individual
sources of uncertainty to the result's covariance matrix may be studied after
unfolding.

\section{Systematic uncertainties on the matrix of migrations}

The matrix of migrations, $\MATRIX{A}$, usually receives uncertainties from various
sources. First, there are statistical uncertainties, originating from
counting the Monte Carlo events in the matrix $\MATRIX{M}$. Second, there may
be systematic uncertainties, in many cases described by a variation
$\MATRIX{M}\to \MATRIX{M}+\MATRIX{\delta M}$, corresponding to a variation of
experimental conditions.

The statistical uncertainties are bin-to-bin independent uncertainties $\Delta
M_{ij}$ on $\MATRIX{M}$. They are propagated through the
unfolding formalism and result in a contribution $\MATRIX{V_{xx}^{M,\text{stat}}}$
to the covariance matrix of \VECTOR{x}. Details are given in the appendix.

A systematic variation $\MATRIX{\delta{M}}$ is propagated to the
result vector in the form of a vector of systematic shifts, 
$\VECTOR{\delta x}$. The corresponding covariance matrix contribution
is given by
$\MATRIX{V_{xx}^{\delta M}}=\VECTOR{\delta x}\TR{(\VECTOR{\delta x})}$. The
calculation of $\VECTOR{\delta x}$ is described in the appendix.
TUnfold supports multiple sources of systematic variation.

\section{Choice of regularisation conditions}
\label{text:regularisation}

Within TUnfold, the matrix of regularisation conditions $\MATRIX{L}$ can be chosen
with some flexibility. Three basic types of regularisation are supported:
\begin{enumerate}
\item rows of \MATRIX{L} where only one element is non-zero,
  corresponding to a regularisation of the amplitude or 
  size of $\VECTOR{x}$,
\item rows of \MATRIX{L} where two elements are non-zero, corresponding to a  
  regularisation of the first derivative of $\VECTOR{x}$,
\item rows of \MATRIX{L} where three elements are non-zero, corresponding to a 
  regularisation of the second derivative (curvature) of $\VECTOR{x}$.
\end{enumerate}
The first derivatives are approximated by differences of event counts in
adjacent bins, $x_{i+1}-x_{i}$. Similarly, the second derivatives are approximated by $(x_{i+1}-x_{i})-(x_{i}-x_{i-1})$.

When initialising TUnfold, it is possible to choose one of the three basic
types of regularisation. This type of regularisation is then applied to all
bins of $\VECTOR{x}$.
\begin{enumerate}
\item if TUnfold is initialised to regularise on the size, $\MATRIX{L}$ is
  initialised to the unity matrix.
\item if TUnfold is initialised to regularise on the first derivatives, 
  $\MATRIX{L}$ has $n-1$ rows and the non-zero elements are: $L_{i,i}=-1$ and
  $L_{i,i+1}=1$. 
\item if TUnfold is initialised to regularise on the second derivatives, 
  $\MATRIX{L}$ has $n-2$ rows and the non-zero elements are: $L_{i,i}=1$, 
  $L_{i,i+1}=-2$, $L_{i,i+2}=1$. 
\end{enumerate}
On the other hand, it is also possible to choose neither of the basic
types and to set up details of the regularisation for specific bins or
groups of bins instead.

\subsection{Regularisation of multi-dimensional distributions}

In many cases, $\VECTOR{x}$ is not simply a one-dimensional distribution. Instead,
the bins of $\VECTOR{x}$ may originate from several distributions,  for example
if there are bins controlling the background normalisation in addition to the
signal bins. Furthermore, the signal bins may originate from a
multi-dimensional distribution. For example, the signal may have $4\times3$
bins in two variables $P_T$ and $\eta$. The vector $x$ then has $12$ bins,
where the first $4$ bins correspond to the $4$ $P_T$ bins of the first $\eta$
bin, etc.
Such a structure is not problematic when regularising on the size, but care
has to be taken when regularising on the first or second derivatives.

Within TUnfold there is support to initialise one-, two- or three-dimensional
regularisation patterns. For example, when regularising the
two-dimensional pattern of $4\times 3$ bins
from the $(P_T,\eta)$ example above on the second derivative, 
$\MATRIX{L}$ is set up as follows:
\begin{equation}
\setcounter{MaxMatrixCols}{12}
\MATRIX{L}= \begin{pmatrix}
  1 & -2 &  1 &  0 &  0 &  0 &  0 &  0 &  0 &  0 &  0 &  0 \\
  0 &  1 & -2 &  1 &  0 &  0 &  0 &  0 &  0 &  0 &  0 &  0 \\
  0 &  0 &  0 &  0 &  1 & -2 &  1 &  0 &  0 &  0 &  0 &  0 \\
  0 &  0 &  0 &  0 &  0 &  1 & -2 &  1 &  0 &  0 &  0 &  0 \\
  0 &  0 &  0 &  0 &  0 &  0 &  0 &  0 &  1 & -2 &  1 &  0 \\
  0 &  0 &  0 &  0 &  0 &  0 &  0 &  0 &  0 &  1 & -2 &  1 \\
  1 &  0 &  0 &  0 & -2 &  0 &  0 &  0 &  1 &  0 &  0 &  0 \\
  0 &  1 &  0 &  0 &  0 & -2 &  0 &  0 &  0 &  1 &  0 &  0 \\
  0 &  0 &  1 &  0 &  0 &  0 & -2 &  0 &  0 &  0 &  1 &  0 \\
  0 &  0 &  0 &  1 &  0 &  0 &  0 & -2 &  0 &  0 &  0 &  1 
 \end{pmatrix}
\end{equation}
Here, rows $1$-$2$ correspond to the regularisation of the second derivatives on
$P_T$ for the first $\eta$ bin. Similarly, rows $3$-$4$ and $5$-$6$ act on $P_T$
for the second and third $\eta$ bin, respectively.
Finally, rows $7$-$10$  correspond to the
second derivatives in $\eta$ for the four $P_T$ bins.

\subsection{Regularisation on the density, multi-dimensional distributions}

The regularisation schemes discussed so far do not take into account the
effects of non-uniform bin widths. 
Another complication arises in cases where multidimensional distributions of
signal and backgrounds have to be mapped to the one-dimensional
vectors $\VECTOR{x}$, $\VECTOR{y}$ and to the matrix $\MATRIX{M}$.
The latest version of TUnfold \cite{tunfold17} addresses 
these issues. Multidimensional distributions 
are mapped on one axis of the vectors 
$\VECTOR{x}$ and $\VECTOR{y}$. The
regularisation conditions may be refined such that the effects of
non-uniform bin widths \cite{harel} are taken into account.

\subsubsection{Densities}
\label{section:density}

During the unfolding, the bins of $\VECTOR{x}$ correspond to event
counts. However, often is desirable to regularise not on the even count but on
the density. The density is calculated
by dividing the number of events in a given bin by the width of the bin.
For calculating the regularisation conditions, the number of events
$\VECTOR{x}$ is transformed to a density $\hat{\VECTOR{x}}$,
\begin{equation}
x_{j} \to \hat{x}_{j}=x_{j}\times\frac{f^{\text{user}}_j}{\prod_d w_{dj}}.
\end{equation}
The number of events $x_j$ is divided by the multi-dimensional bin width $\prod_d
w_{dj}$, where $w_{dj}$ is the bin width of bin $j$ in the 
dimension $d$, as specified by the underlying 
multidimensional distribution. In addition, there is an arbitrary user function
$f^{\text{user}}_j$, which may be used to compensate known kinematic
factors\footnote{An example is the use of the ``reduced cross section'' rather
than the ordinary cross section for inclusive deep-inelastic
scattering \cite{Adloff:1999ah}. The ordinary cross section changes by several orders of
magnitude as a function of kinematic variables and hence is difficult
to regularise. In contrast, the reduced cross section does not vary a
lot, and thus is more natural to regularise on.}.
In TUnfold, the transformation to the density is implemented by
modifying the elements of the matrix $\MATRIX{L}$,
\begin{equation}
L_{rj} \to L_{rj}\times\frac{f^{\text{user}}_j}{\prod_d w_{dj}},
\end{equation}
where the index $r$ is used to enumerate the $n_R$
regularisation conditions.

\subsubsection{Derivatives}

In the case where the regularisation is made on the derivatives, the
bin width may also be included in the approximate calculation of the
derivatives. The calculation of first derivatives is modified such
that
\begin{equation}
(x_{j_2}-x_{j_1}) \to \frac{\Delta_d}{\delta^{d}_{j_2j_1}}(x_{j_2}-x_{j_1}),
\end{equation}
where $j_2$ and $j_1$ are the indices of adjacent bins of a multi-dimensional
distribution and  $d$ is the dimension of the distribution for which
the derivative is calculated. The distance between the two bin centres
is $\delta^{d}_{j_2j_1}$ and $\Delta_d$ is a normalisation constant
specific to the dimension $d$. In TUnfold, the normalisation constant
by default is
chosen to be the average bin width in dimension $d$.
The $\Delta_d$ are relevant if derivatives are considered for
multi-dimensional distributions, where often the derivatives along
one dimension are different in magnitude from derivatives along
another dimension. For example, in the variable $P_T$ the typical bin
width may be $10\,[\text{GeV}]$, where as in $\eta$ the typical bin
width may be $0.5$. In this case, the
derivatives in $P_T$ typically are a factor of $20$ smaller than those in
$\eta$, unless the normalisation $\Delta_d$ is chosen appropriately.

In analogy to the case of first derivatives, the calculation of second
derivatives may be modified to take into account bin widths using the
transformation
\begin{equation}
(x_{j_3}-x_{j_2})-(x_{j_2}-x_{j_1}) \to 
\frac{(\Delta_d)^2}{\delta^{d}_{j_2j_1}+\delta^{d}_{j_3j_2}}\left(\frac{x_{j_3}-x_{j_2}}{\delta^{d}_{j_3j_2}}-\frac{x_{j_2}-x_{j_1}}{\delta^{d}_{j_2j_1}}\right),
\end{equation}
where $j_1$, $j_2$ and $j_3$ are indices corresponding to a triplet of
adjacent bins of a multidimensional distribution.
The distance of bin centres and
normalisation factors are defined similar to the case of first derivatives. 

In TUnfold, the calculation of first or second derivatives including
bin widths is implemented by adding the appropriate modifications to the matrix
$\MATRIX{L}$. It is possible to use the modified calculation of derivatives 
together with the density calculation explained in section
\ref{section:density}.

\subsubsection{Example of a more complicated regularisation scheme}

Consider the use of $4\times 3$ bins in $(P_T,\eta)$, where the bins borders
in $P_t$ are $[5,7,10,15,25]$ and the bin borders in $\eta$ are
$[-2,-0.5,0.5,2]$. The dimension $d=1$ corresponds to $P_T$ and $d=2$
corresponds to $\eta$. In the example, the bin widths along $p_T$ are
$[2,3,5,10]$ and those along $\eta$ are $[1.5,1,1.5]$.
The first four components of the vector $\VECTOR{x}$ hold the four bins in
$P_T$ of the first $\eta$ bin, etc. The bin widths are thus given by
\begin{equation}
\begin{matrix}
w_{1,1}=w_{1,5}=w_{1,9}=2 \\ 
w_{1,2}=w_{1,6}=w_{1,10}=3 \\
w_{1,3}=w_{1,7}=w_{1,11}=5 \\
w_{1,4}=w_{1,8}=w_{1,12}=10
\end{matrix}
\quad\text{and}\quad
\begin{matrix}
w_{2,1}=w_{2,2}=w_{2,3}=w_{2,4}=1.5 \\
w_{2,5}=w_{2,6}=w_{2,7}=w_{2,8}=1\phantom{.5} \\
w_{2,9}=w_{2,10}=w_{2,11}=w_{2,12}=1.5
\end{matrix}
\end{equation}
and average bin sizes are $\Delta_1=5$ and $\Delta_2=1.33$.
The distances of the bin centres are $[2.5,4,7.5]$ along $P_t$ and
$[1.25,1.25]$ along $\eta$, respectively, so
\begin{equation}
\begin{matrix}
\delta^{1}_{2,1}=\delta^{1}_{6,5}=\delta^{1}_{10,9}=2.5 \\
\delta^{1}_{3,2}=\delta^{1}_{7,6}=\delta^{1}_{11,10}=4 \\
\delta^{1}_{4,3}=\delta^{1}_{8,7}=\delta^{1}_{12,11}=7.5
\end{matrix}
\quad\text{and}\quad
\begin{matrix}
\delta^{2}_{5,1}=\delta^{1}_{6,2}=\delta^{2}_{7,3}=\delta^{2}_{8,4}=1.25 \\
\delta^{2}_{9,5}=\delta^{1}_{10,6}=\delta^{2}_{11,7}=\delta^{2}_{12,8}=1.25\,.
\end{matrix}
\end{equation}

The resulting matrix \MATRIX{L} for the case of curvature regularisation on
the density, including bin width effects, then looks like
\begin{equation}
\setcounter{MaxMatrixCols}{12}
\MATRIX{L}= \begin{pmatrix}
  0.51 & -0.56 &  0.13 &  0 &  0 &  0 &  0 &  0 &  0 &  0 &  0 &  0 \\
  0 &  0.12 & -0.11 &  0.02 &  0 &  0 &  0 &  0 &  0 &  0 &  0 &  0 \\
  0 &  0 &  0 &  0 &  0.77 & -0.83 &  0.19 &  0 &  0 &  0 &  0 &  0 \\
  0 &  0 &  0 &  0 &  0 &  0.18 & -0.17 &  0.03 &  0 &  0 &  0 &  0 \\
  0 &  0 &  0 &  0 &  0 &  0 &  0 &  0 &  0.51 & -0.56 &  0.13 &  0 \\
  0 &  0 &  0 &  0 &  0 &  0 &  0 &  0 &  0 &  0.12 & -0.11 &  0.02 \\
  0.19 &  0 &  0 &  0 & -0.57 &  0 &  0 &  0 &  0.19 &  0 &  0 &  0 \\
  0 &  0.13 &  0 &  0 &  0 & -0.38 &  0 &  0 &  0 &  0.13 &  0 &  0 \\
  0 &  0 &  0.08 &  0 &  0 &  0 & -0.22 &  0 &  0 &  0 &  0.08 &  0 \\
  0 &  0 &  0 &  0.04 &  0 &  0 &  0 & -0.11 &  0 &  0 &  0 &  0.04 
 \end{pmatrix}
,
\end{equation}
where the numbers have been rounded to two digits.

\section{Structure of the TUnfold software package}

\begin{figure}[ht]
\begin{center}
\includegraphics[width=0.5\figwidth]{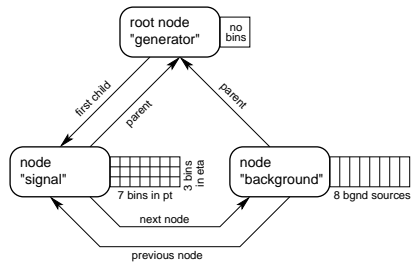}
\caption{\label{fig:binningscheme}example binning scheme with three nodes.
  The ``generator'' node is the root node. It has two child nodes, ``signal''
  and ``background''. The ``signal'' node has a
  two-dimensional binning in two variables, {\sf pt} and {\sf eta}, whereas
  the background node has unconnected bins corresponding to various background
  sources.}
\end{center}
\end{figure}

TUnfold is implemented in the programming language C++ and is interfaced to
the ROOT analysis framework.
The package is organised in four classes
\begin{description}
\item[TUnfold] implements the basic unfolding algorithm and
  L-curve scan.
\item[TUnfoldSys] inherits from the TUnfold class and adds
  functionality to perform background subtraction and propagation of
  systematic uncertainties.
\item[TUnfoldDensity] inherits from the TUnfoldSys class. It adds
  a method to perform scans of global correlations. More important, it provides
  support for multidimensional binning schemes, implemented with the
  help of the class TUnfoldBinning.
\item[TUnfoldBinning] is a class to set up binning schemes. The
  binning schemes are organised in tree-like structures. The nodes 
  of the tree correspond to distinct channels. Each channel may hold a
  multidimensional distribution in some variables. An example of a 
  binning scheme for the vector $\VECTOR{x}$ with signal and background bins is
  shown in figure \ref{fig:binningscheme}.
\end{description}

\section{Summary}
The mathematical foundations of the TUnfold software package have been
presented. TUnfold can be used to correct measurements for migration effects
using the well known mathematical techniques of least-square fitting and
Tikhonov regularisation. 
For choosing the strength of the regularisation parameter, two types of 
scanning methods are implemented: the L curve method and a flexible
minimisation procedure of correlation coefficients. The package offers the
possibility to set up non-trivial regularisation schemes for unfolding
multi-dimensional distribution. Standard methods to subtract background and to
propagate systematic uncertainties are also implemented.

\begin{appendix}

\section{Partial derivatives used for the propagation of uncertainties}

The partial derivatives of $A_{ij}$ with respect to $M_{kj}$ are
\begin{equation}
\frac{\partial A_{ij}}{\partial M_{kj}} = 
\frac{\delta_{ik}-A_{ij}}{s_j}.\label{eqn:partialAM}
\end{equation}
The partial derivatives of $x$ with respect to the
matrix elements $A_{ij}$ are given by
\begin{align}
\frac{\partial x_k}{\partial A_{ij}} = & 
C_{kj} z_i  - (\MATRIX{D^{xy}})_{ki} x_j & \text{where} \\
C_{kj} = & \begin{cases}
E_{kj} & \text{without area constraint} \\
E_{kj}-\frac{(\MATRIX{E}\VECTOR{e})_j(\MATRIX{E}\VECTOR{e})_k}{\VECTORTR{e}\MATRIX{E}\VECTOR{e}} &
\text{with area constraint}
\end{cases} & \text{and} \\
z_i = & \begin{cases}
\left(\MATRIX{V_{yy}}^{-1}(\VECTOR{y}-\MATRIX{A}\VECTOR{x})\right)_i & \text{without area constraint} \\
\left(\MATRIX{V_{yy}}^{-1}(\VECTOR{y}-\MATRIX{A}\VECTOR{x})\right)_i + \frac{\lambda}{2} & \text{with area
    constraint.}
\end{cases} &
\end{align}
In order to derive this result, the partial derivatives of $\MATRIX{E}$
with respect to the elements of the inverse $\MATRIX{E}^{-1}$ are expressed by the
elements of $\MATRIX{E}$, 
\begin{equation}
\frac{\partial E_{ij}}{\partial (\MATRIX{E}^{-1})_{kl}} = -E_{ik}E_{lj}.
\end{equation}
The partial derivative of $x$ with respect to the
regularisation parameter $\tau^2$ is
\begin{equation}
\frac{\partial x_k}{\partial (\tau^2)} = \begin{cases}
\left(\MATRIX{E}(\MATRIXTR{L}\MATRIX{L})(f_b \VECTOR{x_0}-\VECTOR{x})\right)_k & \text{without area constraint} \\
\left(\MATRIX{E}(\MATRIXTR{L}\MATRIX{L})(f_b \VECTOR{x_0}-\VECTOR{x})\right)_k -\frac{\VECTORTR{e}\MATRIX{E}(\MATRIXTR{L}\MATRIX{L})(f_b
    \VECTOR{x_0}-\VECTOR{x})}{\VECTORTR{e}\MATRIX{E}\VECTOR{e}}(\MATRIX{E}\VECTOR{e})_k & \text{with area
    constraint.}
\end{cases}
\end{equation}

\section{Propagation of systematic uncertainties}

Correlated systematic shifts are propagated in the form of systematic
shifts of the result. Given a shift $\MATRIX{\delta M}$ to the matrix
$\MATRIX{M}$, one finds the corresponding shift $\MATRIX{\delta A}$
of $\MATRIX{A}$ using equation \ref{eqn:MtoA}. The resulting shift on
$\VECTOR{x}$ is then given by 
\begin{equation}
%%% PrepareCorrEmat(GetDXDAM(0),GetDXDAM(1),dsys);
%%% PrepareCorrEmat(C,fDXDY,dsys);
%%% *dsysT_VYAx = MultiplyMSparseTranspMSparse(dsys,GetDXDAZ(0));
%%% dsysT_VYAx=dsys^T z
%%% *delta =  MultiplyMSparseMSparse(m1,dsysT_VYAx);
%%% delta = C dsys^T z
%%% *dsys_X = MultiplyMSparseMSparse(dsys,GetDXDAZ(1));
%%% dsys_X = dsys fX
%%% *delta2 = MultiplyMSparseMSparse(m2,dsys_X);
%%% delta2 = fDXDY dsys fX
%%%   AddMSparse(delta,-1.0,delta2);
%%% delta = C dsys^T z - fDXDY dsys fX
% delta = C dsys^T z - fDXDY dsys fX
\VECTOR{\delta x} = \sum_{i,j}\frac{\partial \VECTOR{x}}{\partial
  A_{ij}}(\MATRIX{\delta A})_{ij} = \MATRIX{C}\TR{(\MATRIX{\delta
    A})}\VECTOR{z}-\MATRIX{D^{xy}}(\MATRIX{\delta A})\VECTOR{x},\label{eqn:syserror}
\end{equation}

Statistical uncertainties $\Delta M_{ij}$ of the elements of $\MATRIX{M}$
may also be relevant. The calculation could be done by repeated application of
equations~\ref{eqn:partialAM} and \ref{eqn:syserror} for each independent 
source of uncertainty $\Delta M_{ij}$. However, the required computing costs are large.
In TUnfold, the computation is factorised such
that the computing cost is ${\cal O}(n^3)$
%      // calculate matrices (M1*A)_mj * Z1_j  and  (M1*Rsq)_mj * Z1_j
% *M1A_Z1=MultiplyMSparseMSparse(Dxy,fA);
% ScaleColumnsByVector(M1A_Z1,fX);
%%% *M1Rsq_Z1=MultiplyMSparseMSparse(Dxy,fDAinRelSq);
%%% ScaleColumnsByVector(M1Rsq_Z1,fX);
%%% *AtZ0 = MultiplyMSparseTranspMSparse(fA,z);
%%% *RsqZ0= MultiplyMSparseTranspMSparse(fDAinRelSq,z);
%%% *F=new TMatrixDSparse(*C);
%%% ScaleColumnsByVector(F,AtZ0);
%%% AddMSparse(F,-1.0,M1A_Z1);
%&& *G=new TMatrixDSparse(*C);
%&& ScaleColumnsByVector(G,RsqZ0);
%%% AddMSparse(G,-1.0,M1Rsq_Z1);
%%% r=MultiplyMSparseMSparseTranspVector(F,F,fDAinColRelSq);
%%% *r1=MultiplyMSparseMSparseTranspVector(F,G,0);
%%% *r2=MultiplyMSparseMSparseTranspVector(G,F,0);
%%% Z0sq=z*z
%%% *Z0sqRsq=MultiplyMSparseTranspMSparse(fDAinRelSq,&Z0sq);
%%% *r3=MultiplyMSparseMSparseTranspVector(C,C,Z0sqRsq);
%%% Z1sq = fX*fX
%%% *Z1sqRsq=MultiplyMSparseMSparse(fDAinRelSq,&Z1sq);
%%% *r4=MultiplyMSparseMSparseTranspVector(Dxy,Dxy,Z1sqRsq);
%%% *H=MultiplyMSparseMSparseTranspVector(C,fDAinRelSq,fX);
%%% ScaleColumnsByVector(H,z);
%%% *r5=MultiplyMSparseMSparseTranspVector(Dxy,H,0);
%%% *r6=MultiplyMSparseMSparseTranspVector(H,Dxy,0);
\begin{equation}
\begin{split}(\MATRIX{V_{xx}^{M,\text{stat}}})_{ij} = &
 \sum_k F_{ik} F_{jk} p_k +\sum_k C_{ik}C_{jk} \sum_l Q_{lk}z_l^2 
 +\sum_k D^{xy}_{ik} D^{xy}_{jk} \sum_l Q_{kl}x_l^2 \\
 & - (\MATRIX{F}\MATRIXTR{G}+\MATRIX{G}\MATRIXTR{F})_{ij}
 -(\MATRIX{D^{xy}}\MATRIXTR{H}+\MATRIX{H}\TR{(\MATRIX{D^{xy}})})_{ij}
\quad \text{where}
\end{split}
\end{equation}
\begin{align}
 Q_{ij} = & \left(\frac{\Delta M_{ij}}{s_j}\right)^2 
 \text{ and } p_j = \sum_{i=0}^n Q_{ij}, & \\
 F_{ij} = & \sum_k \frac{\partial x_i}{\partial A_{kj}}A_{kj} = C_{ij}(\MATRIXTR{A}\VECTOR{z})_j - (\MATRIX{D^{xy}A})_{ij} x_j, & \\
 G_{ij} = &  \sum_k \frac{\partial x_i}{\partial A_{kj}}Q_{kj} = C_{ij}(\MATRIXTR{Q}\VECTOR{z})_j - (\MATRIX{D^{xy}}\MATRIX{Q})_{ij} x_j, & \\
 H_{ij} = & z_j \sum_k C_{ik}x_k Q_{jk}. & 
\end{align}

\end{appendix}

\begin{flushleft}

\end{flushleft}
\end{document}